\documentclass[%
superscriptaddress,
 amsmath,amssymb,
 aps,prfluids,
]{revtex4-2}

\usepackage{amsfonts}
\usepackage{amssymb}
\usepackage{amsthm,amsmath}
\usepackage{mathrsfs}
\usepackage{indentfirst}
\usepackage{multirow}
\usepackage{graphicx}
\usepackage{amsmath}
\usepackage[amssymb]{SIunits}
\usepackage{amssymb}
\usepackage{color}
\usepackage[usenames,dvipsnames]{xcolor}
\usepackage{ulem}
\usepackage{cancel}
\usepackage{natbib}
\usepackage[T1]{fontenc}
\usepackage[colorlinks=false, bookmarks=true]{hyperref}
\usepackage{bm}
\newcommand{\dr}[1]{\ifmmode\text{\textcolor{blue}{\sout{\ensuremath{#1}}}}\else\textcolor{blue}{\sout{#1}}\fi}

\begin{document}

\title{\textcolor{black}{Effects of rough walls on sheared annular centrifugal Rayleigh-B\'{e}nard convection}}

\author{Fan Xu}
\affiliation{School of Energy Science and Engineering, Harbin Institute of Technology, Harbin, 150001, P. R. China}
\affiliation{State Key Laboratory of Mesoscience and Engineering, Institute of Process Engineering, Chinese Academy of Sciences, Beijing 100190, P. R. China}
\author{Jun Zhong}
\affiliation{New Cornerstone Science Laboratory, Center for Combustion Energy, Key Laboratory for Thermal Science and Power Engineering of Ministry of Education, Department of Energy and Power Engineering, Tsinghua University, 100084 Beijing, China}
\author{Jinghong Su}
\affiliation{New Cornerstone Science Laboratory, Center for Combustion Energy, Key Laboratory for Thermal Science and Power Engineering of Ministry of Education, Department of Energy and Power Engineering, Tsinghua University, 100084 Beijing, China}
\author{Bidan Zhao}
\affiliation{College of Mechanical and Transportation Engineering, China University of Petroleum, Beijing 102249, P R. China}
\affiliation{State Key Laboratory of Mesoscience and Engineering, Institute of Process Engineering, Chinese Academy of Sciences, Beijing 100190, P. R. China}
\author{Yurong He}
\email{rong@hit.edu.cn}
\affiliation{School of Energy Science and Engineering, Harbin Institute of Technology, Harbin, 150001, P. R. China}
\author{Chao Sun}
\email{chaosun@tsinghua.edu.cn}
\affiliation{New Cornerstone Science Laboratory, Center for Combustion Energy, Key Laboratory for Thermal Science and Power Engineering of Ministry of Education, Department of Energy and Power Engineering, Tsinghua University, 100084 Beijing, China}
\affiliation{Department of Engineering Mechanics, School of Aerospace Engineering, Tsinghua University, Beijing 100084, China}
\author{Junwu Wang}
\email{jwwang@cup.edu.cn}
\affiliation{College of Mechanical and Transportation Engineering, China University of Petroleum, Beijing 102249, P R. China}
\affiliation{State Key Laboratory of Mesoscience and Engineering, Institute of Process Engineering, Chinese Academy of Sciences, Beijing 100190, P. R. China}

\date{\today}

\begin{abstract} 
{
In this study, we investigate the coupling effects of roughness and wall shear in an annular centrifugal Rayleigh-Bénard convection (ACRBC) system, where two cylinders rotate with different angular velocities. Two-dimensional direct numerical simulations are conducted within a Rayleigh number range of $10^{6} \leq Ra \leq 10^{8}$, and the non-dimensional angular velocity difference ($\varOmega$), representing wall shear, varied from 0 to 1. The Prandtl number is fixed at $Pr = 4.3$, the inverse Rossby number at $Ro^{-1} = 20$, and the radius ratio at $\eta = 0.5$. The interaction between wall shear and roughness leads to distinct heat transfer behavior in different regimes. In the buoyancy-dominant regime, an increase in the non-dimensional angular velocity difference ($\varOmega$) significantly enhances heat transfer. However, as $\varOmega$ continues to rise, a sharp reduction in heat transfer is observed in the transitional regime. Beyond a critical value of $\varOmega$, the flow enters a shear-dominant regime, where heat transfer remains unchanged despite further increases in $\varOmega$. The underlying mechanisms behind these distinct heat transfer behaviors are then elucidated. The enhancement of heat transport in the buoyancy-dominant regime is attributed to the introduction of external shear by the rotating rough walls, which reinforces convection and secondary flows within the cavities. Additionally, these secondary flows improve the efficiency of transporting the trapped hot and cold fluids outside the cavities, further enhancing heat transfer. However, in the transitional regime, the number of convection rolls rapidly decreases with increasing shear, eventually reaching a point where sustaining them becomes challenging. Consequently, a sharp reduction in heat transfer occurs. In the shear-dominant regime, heat transfer is primarily governed by diffusion, leading to a constant heat transport rate.
 }
\end{abstract}

\maketitle

\section{Introduction}\label{sec1}
Turbulent thermal convection, a highly complex fluid motion, is widespread in nature and industrial applications. Examples include convection in oceans and the atmosphere, as well as convective flows in heat exchangers and power plant pipes. A classical model extensively used to study turbulent thermal convection is Rayleigh-Bénard convection (RBC) (see \citealp{ahlers2009heat,lohse2010small,chilla2012new} for reviews), where a fluid is confined between two horizontal plates with heating from the bottom and cooling from the top.
In Rayleigh-Bénard (RB) systems, convective heat transport is primarily characterized by the interactions between the large-scale circulation (LSC) and thermal plumes detached from the boundary layers (BLs) \citep{grossmann2000scaling}. However, in many real-life scenarios, thermal convection becomes significantly more complex due to external shear and roughness, both of which influence the characteristics of the LSC and BLs. For example, in atmospheric circulation, the presence of topographical roughness and horizontal winds plays a crucial role in the formation of thermoconvective storms \citep{bluestein2013severe}.

When the RB system is subjected to axial rotation, it undergoes a transition to rotating Rayleigh-Bénard convection (RRBC), as discussed comprehensively in a review by Ecke and Shishkina \cite{ecke2023turbulent}. In RBC, buoyancy serves as the driving force, while in RRBC, centrifugal and Coriolis forces are also introduced. The fundamental focus in studying thermal turbulence lies in understanding the flow dynamics and heat transfer across a wide range of control parameters.
Recently, a novel system called annular centrifugal Rayleigh-Bénard convection (ACRBC) has been proposed for studying Rayleigh-Bénard convection \citep{jiang2020supergravitational,rouhi2021coriolis,wang2022effects,wang2023statistics,zhong2023sheared,zhong2024effect}. This system involves a configuration with cold inner and hot outer cylinders that rotate axially, generating a robust centrifugal force. The enhanced thermal driving force in ACRBC allows for the exploration of higher Rayleigh numbers and facilitates the progression of thermal convection to the ultimate regime \citep{jiang2022experimental}. The observed thermal convection in a rapidly rotating cylindrical annulus serves as an invaluable model for studying flows in planetary cores and stellar interiors \citep{hide1958experimental,busse1974convection,busse1976laboratory,auer1995three,kang2019numerical}, further highlighting the advantages of this system.

The heat transfer and flow dynamics in ACRBC differ from those in classical RBC due to several factors, including the different curvatures of the inner and outer cylinders and the presence of the Coriolis force. Jiang et al. \cite{jiang2020supergravitational} mentioned that when the inverse Rossby number (\textit{Ro}$^{-1}$) is high, following the constraint of the Taylor-Proudman theorem, the flow in ACRBC becomes nearly two-dimensional without axial flow. Besides, through a comparison of two and three-dimensional simulation results, they confirmed that the aspect ratio of ACRBC has minimal impact on heat transfer at high
inverse Rossby numbers (\textit{Ro}$^{-1} \geq 10$). This leads to a reduction in heat transport (\textit{Nu}) compared to the case with low \textit{Ro}$^{-1}$, which measures the influence of the Coriolis force. Furthermore, they observed the formation of four pairs of convection rolls that rotate in the prograde direction around the system's center, known as \textquoteleft{zonal flow}\textquoteright. 
In a subsequent study \cite{jiang2022experimental}, they demonstrated that the ultimate regime in ACRBC occurs when the Rayleigh number $Ra \geq 10^{11}$, which is three orders of magnitude lower than the $Ra \approx 10^{14}$ reported in classical RBC \citep{grossmann2000scaling,he2012transition}. Wang et al. \cite{wang2022effects} focused on investigating the effects of the radius ratio on flow dynamics, heat transport, and temperature distribution in ACRBC. They found that the strength of the zonal flow decreases with increasing radius ratio ($\eta$), while the heat transport efficiency increases with $\eta$. They also observed a deviation of the bulk temperature from the arithmetic mean temperature, with the deviation becoming more significant as $\eta$ decreases. 
Additionally, Zhong et al. \cite{zhong2023sheared} examined the influence of wall shear on heat transfer and flow dynamics in ACRBC. They found that the introduction of shear suppresses heat transfer efficiency and that the flow transitions from turbulent to laminar as the shear increases. Finally, Xu et al. \cite{xu2024effects} observed that the introduction of rough walls enhances heat transport in ACRBC. This enhancement is attributed to the roughness elements promoting the generation and detachment of plumes from the rough walls, thereby strengthening convection in ACRBC. To date, while the individual effects of wall roughness and external shear have been studied in ACRBC, the combined effects of both factors remain unexplored.

In this article, direct numerical simulations (DNS) of ACRBC with shear by the rough walls are carried out to study how the combination of shear and wall roughness affects global heat transport as well as local flow behavior.
The manuscript is organized as follows: In \S 2, the numerical settings are described. In \S 3, the relations between the Nusselt number and the shear strength at different Rayleigh numbers are shown, and the mechanism behind the differences in heat transport is explained. The local flow behavior is also analyzed. Finally, conclusions are drawn in \S 4.

\section{Numerical settings}\label{sec2}
\subsection{Governing equation}\label{sec2.1}
\begin{figure}
	\centering{\includegraphics[width=0.9\textwidth]{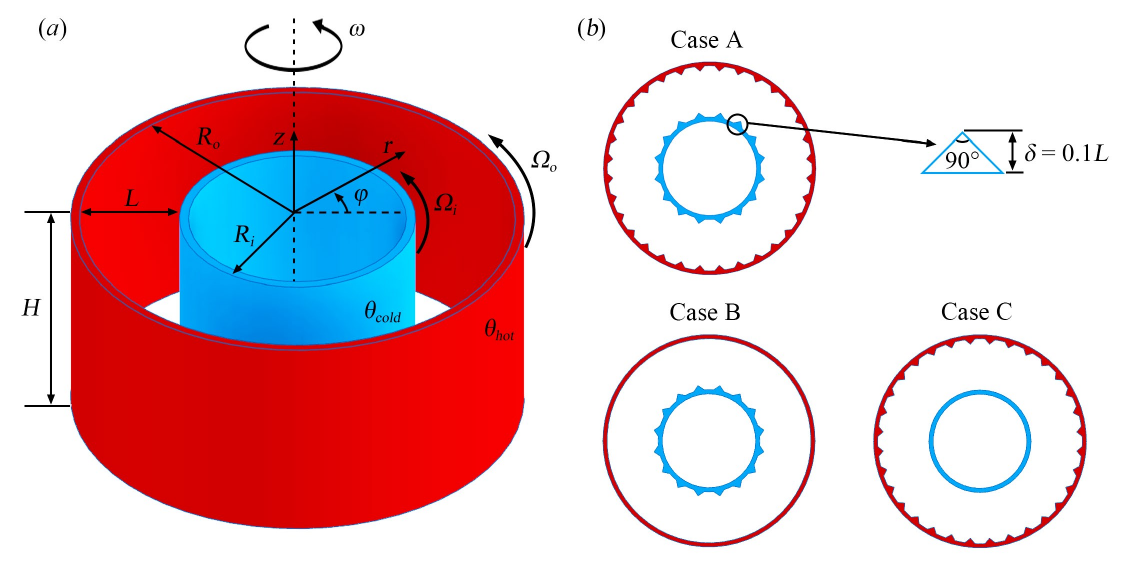}}
	\caption{(\textit{a}) Schematic view of the annular centrifugal Rayleigh-B\'{e}nard convection system and the combination of roughness. (\textit{b}) Three different combinations of roughness, the height of the rough elements is 0.1$L$.}
	\label{ACRBCsystem}
\end{figure}

The study focuses on an annular centrifugal Rayleigh-B\'{e}nard cell bounded by cold inner and hot outer cylinders, which rotate coaxially as shown in figure \ref{ACRBCsystem}(\textit{a}). To show the ACRBC system more clearly, figure \ref{ACRBCsystem}(\textit{a}) is illustrated in a three-dinmentional form. Here, $H$ is the height of the system and $L$ is the gap width. However, the study is conducted by two-dinmentional direct numerical simulations, and the reason is explained later. The system is in a rotating frame with angular velocity $ \omega $, and buoyancy is induced by the centrifugal force ($ \omega $$ r $ + \textit{u}$ _{\varphi} $)$ ^{2} $/\textit{r}, where $ r $ is the radius and $ u_{\varphi} $ is the velocity of the fluid in the azimuthal direction. The motivation of the study is to investigate the influence of the shear force caused by roughness on ACRBC. As illustrated in figure \ref{ACRBCsystem}(\textit{b}), three specific combinations of rough walls are considered, each characterized by a uniform roughness element height of $\delta = 0.1L$. In Case A, the inner cylinder is equipped with sixteen isosceles right triangles that are evenly distributed in the azimuthal direction, while the outer cylinder features thirty-two triangles. In Case B, the same number of roughness elements as in Case A is maintained on the inner cylinder, but the outer wall remains smooth. Conversely, Case C involves a smooth inner cylinder combined with a rough outer cylinder, where the number of roughness elements matches that of Case A.

The motion and heat transfer of the fluid under the Oberbeck-Boussinesq approximation is governed by the non-dimensional Navier-Stokes-Fourier equations in a rotating frame \citep{zhong2023sheared}:
\begin{equation}\label{eq:mass}
	\nabla\bm\cdot\textbf{\textit{u}}=0,
\end{equation}
\begin{equation}\label{eq:momentum}
	\frac{\partial\textit{\textbf{u}}}{\partial\textit{t}}+\nabla\bm\cdot(\textbf{\textit{uu}})=-\nabla\textit{p}-Ro^{-1}\textbf{\textit{e}}_{z}\times\textbf{\textit{u}}+\sqrt{\frac{Pr}{Ra}}\nabla^{2}\textbf{\textit{u}}-\theta\frac{2(1-\eta)}{1+\eta}\bigg(1+\frac{2u_{\varphi}}{Ro^{-1}r}\bigg)^{2}\textit{\textbf{r}},
\end{equation}
\begin{equation}\label{eq:energy}
	\frac{\partial\theta}{\partial\textit{t}}+\nabla\bm\cdot(\textbf{\textit{u}}\theta)=\frac{1}{\sqrt{RaPr}}\nabla^{2}\theta,
\end{equation}
where $ \textbf{\textit{u}} = (\textit{u}_{r}, \textit{u}_{\varphi}, \textit{u}_{z}) $ is the velocity vector, \textit{t} is the time, \textit{p} is the pressure,  \textbf{\textit{e}}$ _{z} $ is the unit vector along the axial direction, $ \theta $ is the temperature, and $ \eta $ is the radius ratio. The equations are normalized using the gap width (without roughness) \textit{L} = \textit{R}$ _{o} $ $ - $ \textit{R}$ _{i} $ for length, the temperature difference between the hot outer cylinder and the cold inner cylinder $ \Delta $ = $ \theta_{hot} $ $ - $ $ \theta_{cold}$ for temperature, the free-fall velocity \textbf{\textit{U}} = $\sqrt{\omega^{2}((R_{i}+R_{o})/2)\alpha\Delta\textit{L}}$ for velocity, and \textit{L}/\textit{U} for time. Here, $ \omega $ donates the angular velocity of the system, and $ \alpha $ is the coefficient of thermal expansion of the fluid. In our system, considering both physical meaning and simplicity, we set $ \omega = (\varOmega_{i} + \varOmega_{o})/2 $. This choice provides a good estimate for the free-fall velocity when the shearing is relatively small, implying that $ \varOmega_{i} - \varOmega_{o} \ll \omega $. When the shearing becomes strong enough and dominates the flow, the contribution of buoyancy becomes insignificant and can be neglected, as will be discussed later. Therefore, the selected $ \omega $ is reasonable in most of the parameter space. Under this rotating frame, the inner cylinder rotates at a non-dimensional angular velocity $ \varOmega = (\varOmega_{i} - \omega)L/U $, while the outer cylinder rotates at $ -\varOmega $. Additionally, the choice of this non-dimensional angular velocity difference aligns with the approach taken by Zhong et al. \cite{zhong2023sheared}, facilitating direct comparison of results with theirs. In the coordinate system \textit{r}, $ \varphi $ , \textit{z} refer to the wall-normal (radial), streamwise (azimuthal) and spanwise (axial) directions. In the following, we define the non-dimensional radius \textit{R}$ ^{\ast} $ to be \textit{R}$ ^{\ast} $ = (\textit{r} $ - $ \textit{R}$ _{i} $)/(\textit{R}$ _{o} $ $ - $ \textit{R}$ _{i} $).

From the above non-dimensional equations, it can be seen that there are five dimensionless control parameters in ACRBC. The Rayleigh number (buoyancy-driven strength)
\begin{equation}\label{eq:Ra}
	Ra=\omega^{2}(R_{i}+R_{o})\alpha\Delta{L}^{3}/(2\nu\kappa),
\end{equation}
and the Prandtl number (fluid physical property)
\begin{equation}\label{eq:Pr}
	Pr = \nu/\kappa,
\end{equation}
as in classical RBC, where $ \nu $ and $ \kappa $ are the kinematic viscosity and thermal diffusivity of the fluid, respectively. Three additional control parameters are the inverse Rossby number
\begin{equation}\label{eq:1overRo}
	Ro^{-1} = 2\omega{L}/U,
\end{equation}
which measures the effects of Coriolis force, the radius ratio that measures the geometric property
\begin{equation}\label{eq:eta}
	\eta= \textit{R}_{i}/\textit{R}_{o},
\end{equation}
and the angular velocity difference of two cylinders
\begin{equation}\label{eq:omega}
	\varOmega = (\varOmega_{i} - \omega)L/U.
\end{equation}
In addition, the key response parameter is the Nusselt number measuring the efficiency of heat transport \citep{jiang2020supergravitational,wang2022effects}
\begin{equation}\label{eq:Nu}
	Nu=J/J_{con}=(\langle\textit{u}_{r}\theta\rangle _{\varphi,z,t}-\kappa\partial_{r}\langle\theta\rangle_{\varphi,z,t})/(\kappa\Delta(r{\cdot}ln(\eta))^{-1}),
\end{equation}
where $ J, J_{con}, u_{r}$, and $ \theta $ are the total heat flux, the heat flux through pure thermal conduction, the radial velocity and the temperature of a certain point, respectively. Here, $ \langle...\rangle $$_{\varphi,z,t}$ denotes averaging over a cylindrical surface (averaging over the axial and azimuthal directions) with constant distance from the axis and over time. Note that the expression of conductive heat flux in cylindrical geometry is slightly different from that in classical RBC, the detailed derivation process can be found in our previous study \citep{wang2022effects}.

\subsection{Direct numerical simulations}\label{sec2.2}
Equations ($ \ref{eq:mass} $) $ - $ ($ \ref{eq:energy} $) are solved using a second-order-accuracy, colocated finite-volume method in the Cartesian coordinate system, using OpenFOAM as the computational platform. The rough elements are dealt with a second-order-accuracy immersed boundary method \citep{zhao2020cfd,zhao2020Acomputational}. Based on the findings of Jiang et al. \cite{jiang2020supergravitational}, the flow exhibits near-two-dimensionality (without axial flow) at high inverse Rossby numbers \textit{Ro$ ^{-1} $} ($ \geq $ 10), as dictated by the Taylor-Proudman theorem. Therefore, in this study, two-dimensional direct numerical simulations (2D-DNS) are conducted, maintaining a fixed inverse Rossby number of $ Ro^{-1} = 20 $. No-slip boundary condition is used for velocity and constant temperature boundary condition is used for the temperature at two cylinder walls.

\begin{figure}
	\centering{\includegraphics[width=0.9\textwidth]{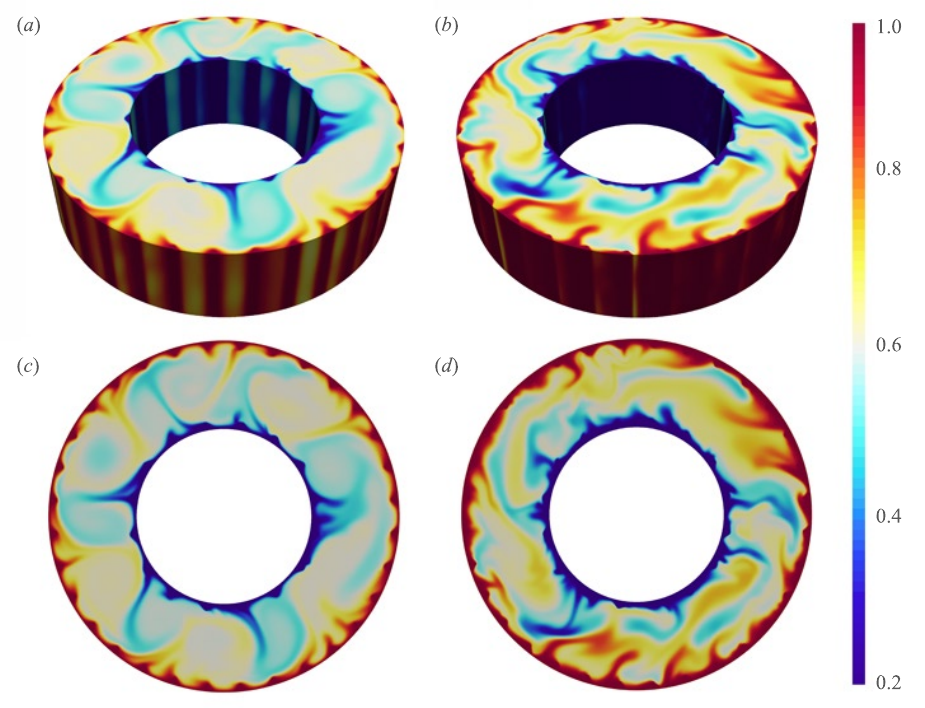}}
	\caption{The instantaneous temperature fields for Case A obtained from the three-dimensional simulation at $\textit{Ra}=10^6$, shown at (\textit{a}) $\varOmega=0$ and (\textit{b}) $\varOmega=0.1$. (\textit{c}) and (\textit{d}) show the instantaneous temperature fields at the middle planes of (\textit{a}) and (\textit{b}), respectively.}
	\label{validation}
\end{figure}

In order to check the reliability of the 2D simulations in ACRBC system with rough walls, we conduct a validation of the two-dimensional model by using three-dimensional simulation for $Ra=10^6$ at $\varOmega=0$ and $\varOmega=0.1$ (representing the cases with and without shearing). As shown in figure \ref{validation}, the instantaneous temperature fields at each $r-\varphi$ plane are nearly identical, regardless of the presence of shear. This indicates that even with rough inner and outer walls, the axial flow is significantly restricted due to the strong Coriolis force ($ Ro^{-1} = 20$) in the ACRBC. This result is consistent with the findings of Jiang et al. \cite{jiang2020supergravitational} in ACRBC and Zhong et al. \cite{zhong2023sheared} in sheared ACRBC, where similar effects were observed under smooth-wall conditions within the same $Ra$ range (from $Ra = 10^6$ to $Ra = 10^8$) as in our study. It is important to note that with the increase in Rayleigh number, the Coriolis force's suppressive effect on axial flow weakens \cite{jiang2020supergravitational}. This occurs because the stronger buoyancy-driven convection (at higher $Ra$) competes with and partially overcomes the effects of the Coriolis force, leading to a more efficient axial flow in the system. 
Although the dimensionality of the simulation may affect heat transport in ACRBC as $Ra$ increases, our comparisons show that the key features of convection, including temperature and velocity fields, are well-represented in the two-dimensional approximation for the parameters used in our study. However, we acknowledge that while the two-dimensional model captures the basic convection patterns, it may not fully capture the complex three-dimensional flow structures that are essential for understanding heat transfer in sheared ACRBC at high $Ra$. As the interaction between shear, rotation, and buoyancy becomes more significant for accurate heat transfer predictions, future work must focus on three-dimensional simulations to more effectively capture these interactions.

To ensure adequate resolutions, we carefully examined the spatial and temporal resolutions for all simulations, aiming to capture all relevant scales accurately. The ratios of the maximum grid spacing, $ \Delta_{g} $, to the Kolmogorov scale estimated by the global criterion $ \eta_{K}=LPr^{1/2}/[Ra(Nu-1)]^{1/4}\cdot[(1+\eta)ln(\eta)/2(\eta-1)]^{1/4} $ \citep{jiang2020supergravitational} is maintained below 1.0 ($ \Delta_{g}/\eta_{K}<1.0 $). Additionally, the ratio of $ \Delta_{g} $ to the Batchelor scale, $ \eta_{B}=\eta_{K}Pr^{-1/2} $ \citep{silano2010numerical}, is kept below 2.0 ($ \Delta_{g}/\eta_{B}<2.0 $).
Moreover, the grid is uniform in the azimuthal direction, while it is refined near the inner and outer cylindrical walls in the radial direction to ensure appropriate spatial resolution within the boundary layers (BLs). Specifically, there are a minimum of 8 grid points inside the thermal BLs and 10 grid points inside the viscous BLs.
For temporal discretization, the second-order backward scheme is employed for the temporal term, while a second-order total variation diminishing (Vanleer) scheme is used for the convective term. The simulations are conducted with a fixed time step based on the Courant-Friedrichs-Lewy (CFL) criterion, ensuring that the CFL number remains below 0.7 in all simulations.
To achieve statistical convergence, the simulations are run for a sufficient duration (80 free-fall time) after the system reaches a statistically stationary state (100 free-fall time). The relative difference of \textit{Nu} based on the first and second halves ($ \epsilon_{Nu} = |(\langle$\textit{Nu}$\rangle_{0-T/2} - \langle$\textit{Nu}$\rangle_{T/2-T})|/\textit{Nu} $) of the simulations is less than 1 \%. All relevant details are presented in table (\ref{table1})$ - $(\ref{table3}) in Appendix \ref{appendix}. As mentioned above, the simulations are conducted in Cartesian coordinates. However, the ACRBC system is more naturally described in cylindrical coordinates, and the results are subsequently transformed into this format during the simulations. To ensure the expression aligns better with the conventions of ACRBC, the resolutions in the Appendix are presented as $N_r$$\times$$N_\varphi$ instead of $N_x$$\times$$N_y$.

The primary objective of this study is to investigate the influence of roughness on the flow dynamics and heat transport in the sheared ACRBC system. Among the various options for roughness shapes, we specifically focus on the isosceles right triangular rib as the simplest geometric model. This choice allows us to analyze its impact on the statistical properties of turbulent sheared ACRBC. Our main research question involves understanding how the presence of roughness elements affects both the global transport and local flow statistics in sheared ACRBC turbulence. To facilitate direct comparisons with previous findings \citep{zhong2023sheared}, we set the radius ratio, denoted as \textit{$\eta$}, to 0.5. In all simulations, we maintain a fixed Prandtl number (\textit{Pr}) of 4.3, corresponding to the working fluid properties of water at a temperature of 40$^{\circ}$C. As mentioned earlier, the inverse Rossby number is set to $ Ro^{-1} = 20$, enabling us to explore a range of Rayleigh numbers (\textit{Ra}) from $10^6$ to $10^8$ and a range of angular velocity differences $ \varOmega $ from 0 to 1 using the 2D-DNS method. A positive value of $ \varOmega $ indicates that the inner cylinder rotates at a higher angular velocity than the outer cylinder in the stationary reference frame. Detailed simulation parameters can be found in table (\ref{table1})$ - $(\ref{table3}) in Appendix \ref{appendix}.

\section{Results and discussion}
\subsection{Heat transport and boundary layer}

\begin{figure}
	\centering
	{\includegraphics[width=1\textwidth]{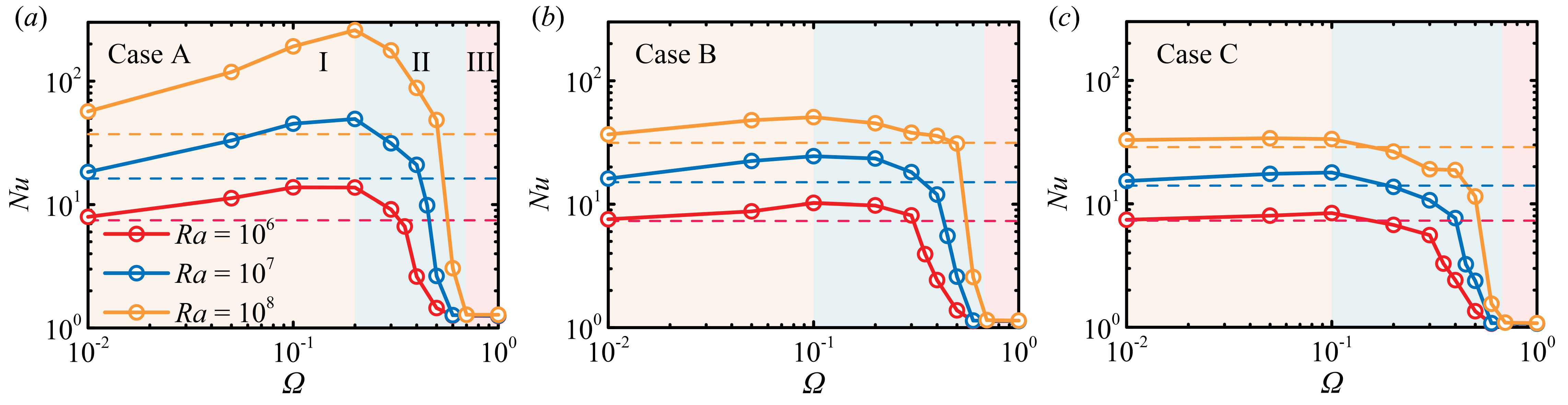}}
	\caption{Nusselt number \textit{Nu} as a function of non-dimensional angular velocity difference $\varOmega$ for (\textit{a}) Case A, (\textit{b}) Case B, and (\textit{c}) Case C at three different Rayleigh numbers . The dashed lines of the corresponding colors for each \textit{Ra} indicate the value of \textit{Nu} at $\varOmega = 0$.}\label{NuvsOmega}
\end{figure}

We commence our analysis by investigating the influence of shear force exerted by rough walls on heat transport in ACRBC. Research conducted with smooth walls in ACRBC \citep{zhong2023sheared,zhong2024effect}, Couette-RBC \citep{blass2020flow,blass2021effect}, and Poiseuille-RBC \citep{scagliarini2014heat} demonstrated an initial suppression of heat transport due to shear-induced effects. However, the scenario changes when the walls are not smooth. As illustrated in figure \ref{NuvsOmega}, the Nusselt number (\textit{Nu}) is plotted against the non-dimensional angular velocity difference ($\varOmega$) for three distinct combinations of roughness elements (Cases A, B, and C) at $Ra = 10^6$, $Ra = 10^7$, and $Ra = 10^8$.
For each case, as $\varOmega$ increases, three universal regimes emerge, distinguished by different background colors. Additionally, two critical non-dimensional angular velocity differences, $\varOmega_{c1}$ and $\varOmega_{c2}$, are identified for each \textit{Ra}. However, there are variations in the values of $\varOmega_{c1}$ among the cases. Specifically, the value of $\varOmega_{c1}$ for Case A is smaller compared to that for Cases B and C. This dissimilarity can be attributed to the fact that both inner and outer walls are rough in Case A, whereas Case B (Case C) only features a rough inner (outer) wall, with the other wall being smooth.

Now, let's focus on Case A (figure \ref{NuvsOmega}\textit{a}). It can be observed that in the regime below $\varOmega_{c1}$, heat transport experiences a significant enhancement caused by wall shear, referred to as regime I, the buoyancy-dominant regime. Conversely, above $\varOmega_{c2}$, a nearly purely diffusive state is observed, resulting in the minimum value of \textit{Nu}. This is referred to as regime III, the shear-dominant regime. Between these two regimes, \textit{Nu} experiences a sharp decrease, indicating an abrupt transition referred to as regime II (the transitional regime).

The regime behaviors in Case B and Case C resemble those of Case A, although they are less pronounced. Within each case, in Regimes I and II, the heat transfer efficiency (\textit{Nu}) increases with increasing \textit{Ra}. However, as the shear force intensifies and the flow becomes more stable, the values of \textit{Nu} for different \textit{Ra} converge in Regime III, as shown in figure \ref{NuvsOmega}. Notably, the critical non-dimensional angular velocity difference $\varOmega_{c2}$ for $Ra = 10^8$ is slightly larger than the values for $Ra = 10^6$ and $Ra = 10^7$ (changing from $\varOmega_{c2} = 0.7$ to 0.6). This can be attributed to the fact that a stronger shear force is required to transition the flow towards a laminar state under strong centrifugal buoyancy.
It is worth mentioning that the Nusselt number (\textit{Nu}) for the rough cases in Regime III exceeds that for the smooth case. This outcome can be attributed to the slight disturbance caused by wall roughness, preventing the flow from reaching a completely laminar state. Consequently, the \textit{Nu} value for the rough case exceeds 1, which is the \textit{Nu} value for the smooth case in the laminar regime. Additionally, compared to a smooth wall, a rough wall has a larger surface area due to the presence of rough structures, which increases the effective contact area with the fluid. Even in the absence of radial velocity, the heat exchange area between the wall and the fluid is enlarged.

\begin{figure}
	\centering
	{\includegraphics[width=1\textwidth]{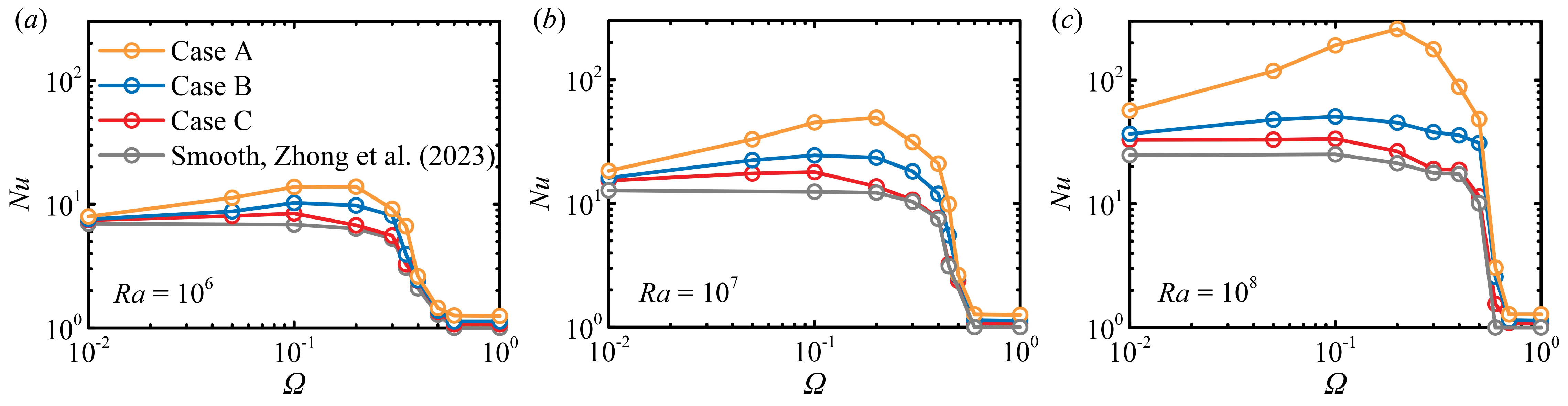}}
	\caption{Nusselt number \textit{Nu} as a function of non-dimensional angular velocity difference $\varOmega$ for three different rough cases at (\textit{a}) $Ra = 10^6$, (\textit{b}) $Ra = 10^7$, and (\textit{c}) $Ra = 10^8$. The corresponding results for the smooth case \citep{zhong2023sheared} at the same Rayleigh numbers are also presented for comparison.}\label{NuvsOmega_1}
\end{figure}

To demonstrate the effects of enhancement, we include three dashed lines in figure \ref{NuvsOmega}: the red dashed line represents the value of $Nu$ at $\varOmega = 0$ for $Ra = 10^6$, the blue dashed line represents the value of $Nu$ at $\varOmega = 0$ for $Ra = 10^7$, and the orange dashed line represents the value of $Nu$ at $\varOmega = 0$ for $Ra = 10^8$. For the purpose of analysis, we focus on Case A, which exhibits the most significant heat enhancement among the considered scenarios. As shown in figure \ref{NuvsOmega}(\textit{a}), it can be observed that a maximal enhancement of 5.79 times is achieved at $\varOmega = 0.2$ for $Ra = 10^8$. This significant increase in \textit{Nu} is noteworthy as it qualitatively deviates from the findings of Zhong et al. \cite{zhong2023sheared} in smooth ACRBC, where \textit{Nu} initially decreases with $\varOmega$ before reaching a minimum (see figure \ref{NuvsOmega_1}). It is important to note that Wagner et al. \cite{wagner2015heat} established an upper bound on heat transfer enhancement solely due to regular surface roughness, expressed as $Nu/Nu_{s}(0) \leq A_{w}/A_{s}$, where $Nu_{s}(0)$ represents the Nusselt number in smooth cases without shear, $A_{w}$ is the wetted covering surface area over the roughness, and $A_{s}$ is the covering area over a smooth surface. However, our study reveals that the maximal enhancement rate $Nu/Nu_{s}(0) \approx 5.79$ at $Ra = 10^8$ and $\varOmega = 0.2$ far exceeds the upper bound $A_{w}/A_{s} \approx 1.72$ under the influence of external shear. In other words, the enhancement of heat transport is primarily influenced by the shear of rough walls rather than solely by the increase of surface area. This enhancement of heat transport through shearing of rough walls has also been reported by Jin et al. \cite{jin2022shear} in classical Rayleigh-B\'{e}nard Convection (RBC). The mechanism behind how the shear of rough walls enhances heat transfer will be discussed later.

Figure \ref{NuvsOmega_1} illustrates the relationship between the Nusselt number \textit{Nu} and the non-dimensional angular velocity difference $\varOmega$ for three different rough cases with $Ra = 10^6$, $Ra = 10^7$, and $Ra = 10^8$. The corresponding results for the smooth case \citep{zhong2023sheared} at the same Rayleigh numbers are also presented for comparison. The figure reveals that although the trends of \textit{Nu} varying with $\varOmega$ are consistent for different cases at the same $Ra$, there are some distinctions among them. At each Rayleigh number, the values of \textit{Nu} follow a descending order (Case A $>$ Case B $>$ Case C) at the same $\varOmega$. The highest Nusselt number (\textit{Nu}) observed in Case A at the same $\varOmega$ can be attributed to the roughness of both the inner and outer cylinders in this case. In Case B, the inner cylinder is rough with sixteen isosceles right triangles, while the outer cylinder is smooth. Conversely, in Case C, the inner cylinder is smooth, and the outer cylinder is rough with thirty-two isosceles right triangles. Despite Case B having a lower number of roughness elements compared to Case C, the \textit{Nu} of Case B surpasses that of Case C. This disparity can be explained as follows: In regimes I and II, heat transport primarily depends on the convection of hot plumes detached from the outer wall and cold plumes detached from the inner wall. In Case B, the detached cold plumes from the inner rough wall collide with the concave surface of the smooth outer wall, leading to effective \textquoteleft{rebounding}\textquoteright by the outer wall. However, in Case C, the detached hot plumes from the outer rough wall encounter the convex surface of the smooth inner wall, which limits their effective \textquoteleft{rebounding}\textquoteright. As a result, despite Case B having fewer roughness elements compared to Case C, the \textit{Nu} of Case B is greater than that of Case C. In regime III, characterized by nearly laminar flow, heat transfer primarily relies on thermal conduction. According to equation ($ \ref{eq:Nu} $), the value of \textit{Nu} at this stage increases with an increasing radius ratio $ \eta $. The effective radius ratio for Case A is approximately 0.58, for Case B it is approximately 0.55, and for Case C it is approximately 0.53. Consequently, the \textit{Nu} values exhibit a descending order (Case A $>$ Case B $>$ Case C) in regime III, as depicted in figure \ref{NuvsOmega_1}.

Based on the analysis conducted above, it is discovered that the coupling of wall shear and roughness in ACRBC can lead to a significant enhancement in heat transfer efficiency under specific conditions. Particularly, when the shear strength is relatively low, buoyancy dominates the heat transfer process, resulting in a substantial improvement. Surprisingly, at certain combinations of \textit{Ra} and $\varOmega$, the heat transfer efficiency can reach its maximum, surpassing the previously established upper limit of heat transfer enhancement. However, as the shear strength exceeds a certain threshold, the flow transitions into a diffusive regime, resulting in the minimum value of \textit{Nu}.Additionally, the presence of rough walls, which increases the effective contact area and introduces slight disturbances, prevents the flow from entering a laminar state. Moreover, different combinations of rough walls have varying effects on \textit{Nu}, with the highest \textit{Nu} observed when both the inner and outer walls are rough, followed by the case where only the inner wall is rough while the outer wall is smooth, and finally, the case where the outer wall is rough while the inner wall is smooth.

\begin{figure}
	\centering
	{\includegraphics[width=0.5\textwidth]{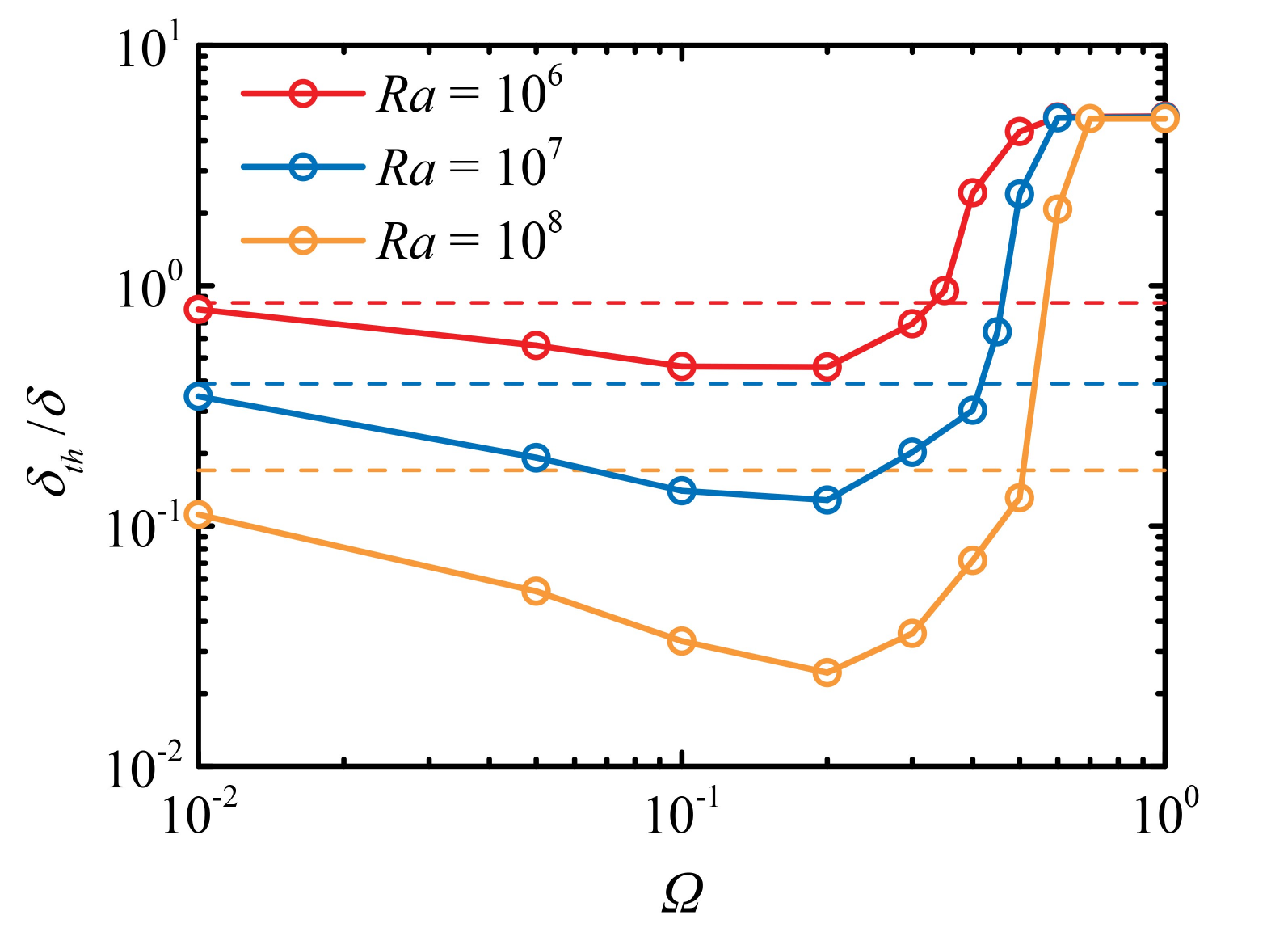}}
	\caption{Dimensionless value $ \delta_{th}/ \delta $ of the roughness height as a function of the Rayleigh number for Case A. Here, $ \delta_{th} $ donates the estimated thickness of the thermal boundary layer, given by $\delta_{th} \approx \sigma d / (2Nu)$, where $ \sigma $ is the geometry factor defined as $ \sigma = [(R_{i}+R_{o})/(2\sqrt{R_{i}R_{o}})]^{4} $ \citep{brauckmann2013direct}, and $ \delta $ is the height of roughness element. The dashed lines of the corresponding colors for each \textit{Ra} indicate the value of $ \delta_{th}/ \delta $ at $\varOmega = 0$.}\label{BL}
\end{figure}

It is commonly observed that roughness can enhance heat transfer within the conduction layer when the roughness height is comparable to or less than the thickness of the thermal boundary layer, showing with an increase in prefactor or scaling law \citep{shen1996turbulent,stringano2006turbulent,jiang2018controlling}. However, as the Rayleigh number continues to increase and the roughness height becomes much larger than the thermal boundary layer thickness, the classical scaling law $Nu \sim Ra^{1/3}$ should be observed rather than the ultimate scaling of $Nu \sim Ra^{1/2}$, since in this regime, the vortices generated by the roughness are no longer confined within the conduction layer \citep{zhu2017roughness}. Figure \ref{BL} vividly presents the normalized thermal BL thickness, $\delta_{th}/\delta$, as a function of shear strength ($\varOmega$) for various Rayleigh numbers in Case A. Here, the thermal boundary layer thickness, $\delta_{th}$, is estimated using the formula $\delta_{th} \approx \sigma d / (2Nu)$, in which $\sigma$ represents the geometry factor defined as $\sigma = [(R_i + R_o)/(2\sqrt{R_i R_o})]^4$ \citep{brauckmann2013direct},
and the red, blue and orange dashed lines represent respectively the values of $ \delta_{th}/ \delta $ at $\varOmega = 0$ for $Ra = 10^6$, $Ra = 10^7$ and $Ra = 10^8$.
Noting that figure \ref{BL} is intended to offer an alternative view of the data presented in figure \ref{NuvsOmega}($a$).
As depicted in figure \ref{BL}, the non-dimensional thermal BL thickness, $\delta_{th}/\delta$, for the three different Rayleigh numbers is initially less than 1 at the beginning of the explored parameter range ($\varOmega = 0.01$), indicating that the thermal BL thickness is smaller than the height of the roughness elements at this $\varOmega$. With increasing shear, $\delta_{th}/\delta$ initially decreases and then gradually increases in regime II until reaching a maximum value in regime III. This opposite trend compared to that of \textit{Nu} indicates a distinct changing process. It is worth noting that in regime II, the thermal boundary layer thicknesses for different Rayleigh numbers become larger than the height of the roughness elements as the shear of the rough walls strengthens. This observation suggests a diminishing influence of roughness on heat transfer with increasing shear within this regime.
Furthermore, figure \ref{BL} reveals that the thickness of the thermal boundary layer follows a descending order for $Ra = 10^6$, $Ra = 10^7$, and $Ra = 10^8$ at the same $\varOmega$ in regime I and II. This indicates that the impact of the roughness elements on flow dynamics and heat transfer is most pronounced in $Ra = 10^8$, followed by $Ra = 10^7$, and least significant in $Ra = 10^6$ in these two regimes. However, with further increasing $\varOmega$, the thermal boundary layer thicknesses for these three Rayleigh numbers become equal in regime III. Consequently, the \textit{Nu} values for each $Ra$ become identical. The findings presented in figure \ref{NuvsOmega} are strongly supported by these quantitative results.

\subsection{The mechanism of heat transport enhancement}

As mentioned earlier, the enhancement of Nusselt number (\textit{Nu}) is observed in the buoyancy-dominant regime of ACRBC with rough walls, as depicted in figure \ref{NuvsOmega}. This finding is in stark contrast to previous studies conducted with smooth walls in ACRBC \citep{zhong2023sheared,zhong2024effect}, Couette-RBC \citep{blass2020flow,blass2021effect}, and Poiseuille-RBC \citep{scagliarini2014heat}, where \textit{Nu} were initially suppressed due to shear-induced effects. Therefore, a crucial question arises: how is the heat transfer efficiency improved in rough wall cases of ACRBC?

\begin{figure}
	\centering
	{\includegraphics[width=0.95\textwidth]{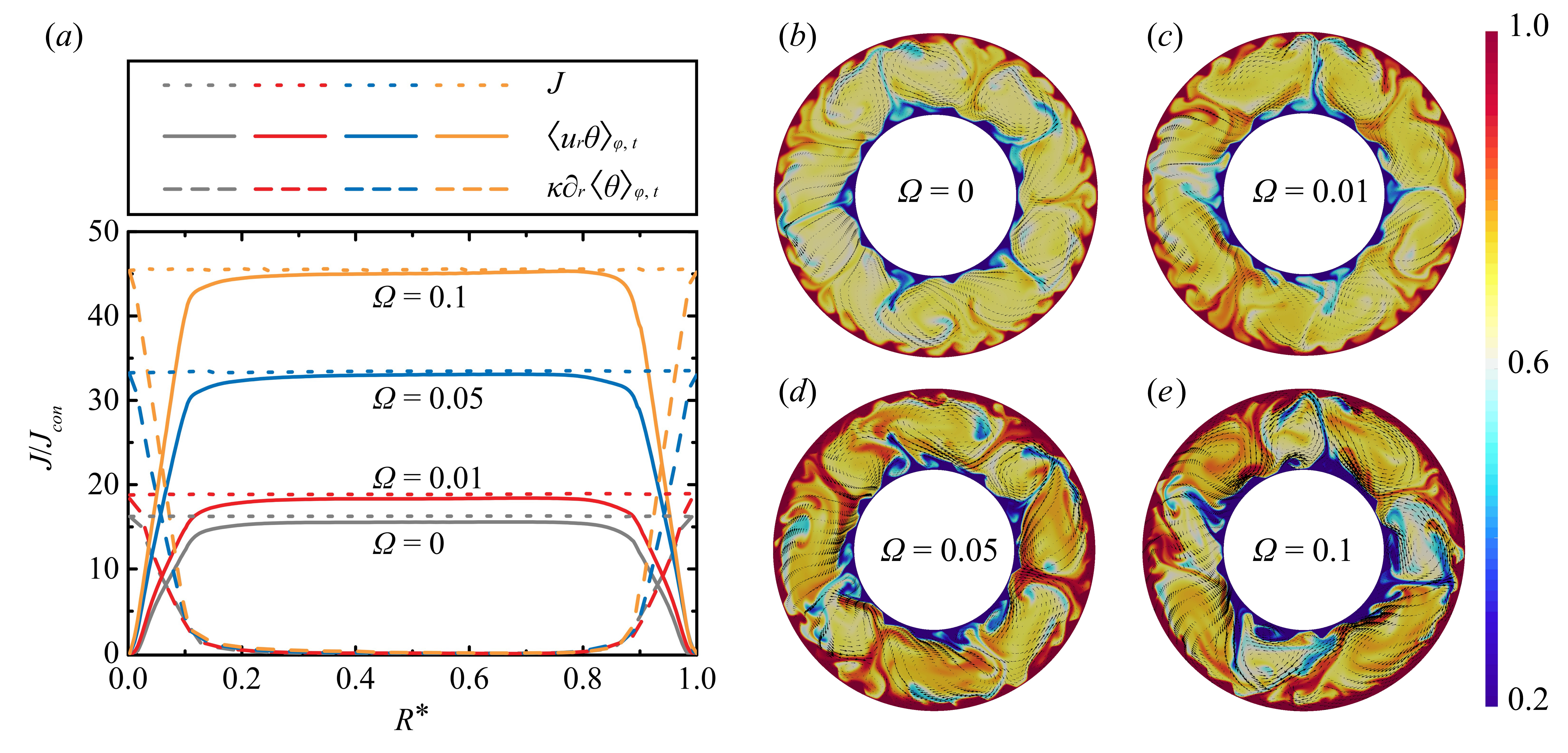}}
	\caption{The convective and diffusive contributions to the heat flux for Case A at different non-dimensional angular velocity difference $ \varOmega $ and \textit{Ra} = 10$ ^{7} $ (\textit{a}), where all results are normalized by the heat flux of the non-vortical laminar state \textit{J}$ _{con}$. The corresponding snapshots of temperature field superposed with instantaneous velocity vectors in cases of $\varOmega = 0$ (\textit{b}), $\varOmega = 0.01$ (\textit{c}), $\varOmega = 0.05$ (\textit{d}) and $\varOmega = 0.1$ (\textit{e}).}\label{decompose_Nu}
\end{figure}

\begin{figure}
	\centering
	{\includegraphics[width=0.5\textwidth]{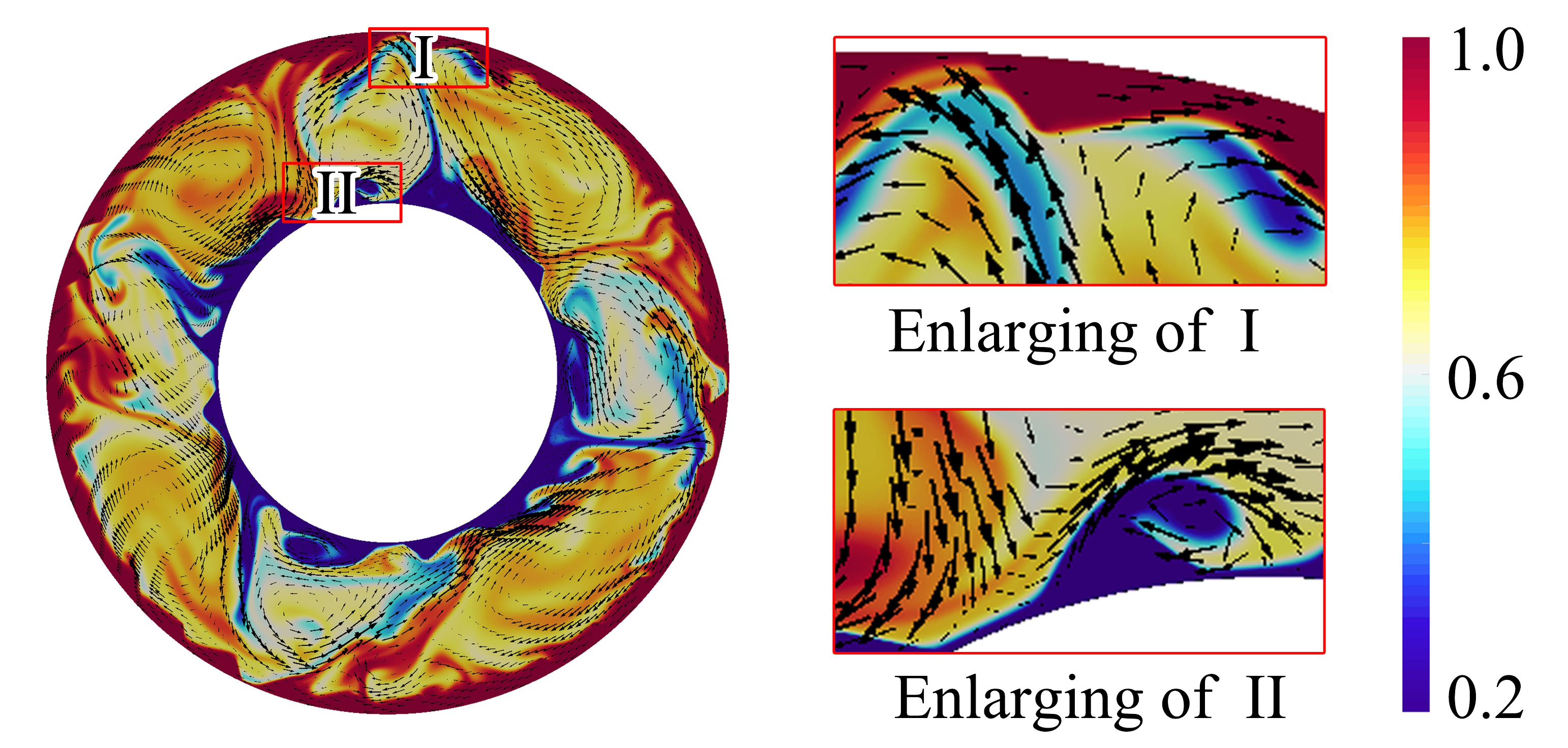}}
	\caption{The temperature field superposed with instantaneous velocity vectors and its local enlarged drawings for Case A at \textit{Ra} = 10$ ^{7} $ and $\varOmega = 0.1$.}\label{enlarge}
\end{figure}

In Rayleigh-B\'{e}nard convection, the heat flux is calculated as $ J = \langle\textit{u}_{r}\theta\rangle _{\varphi,t}-\kappa\partial_{r}\langle\theta\rangle_{\varphi,t} $. Here, the first term represents the convective contribution, and the second term represents the diffusive contribution \citep{wang2022effects}. Figure \ref{decompose_Nu}(\textit{a}) illustrates the radial profiles of these two contributions for Case A at different shear strengths ($\varOmega$) and $Ra = 10^7$. The dotted, solid and dashed lines respectively correspond to the total heat flux, the convective and diffusive contributions. The gray, red, blue and orange lines respectively correspond to $\varOmega = 0$, $\varOmega = 0.01$, $\varOmega = 0.05$ and $\varOmega = 0.1$. It is worth noting that the total heat flux shows minimal variation across the radius, indicating sufficient spatial and temporal resolutions in our simulations.
Furthermore, it can be observed that the convective contribution to the heat flux is predominantly concentrated in the central region and diminishes towards the boundaries, as expected. In contrast, the diffusive contribution dominates near the walls but diminishes to nearly zero in the middle. More importantly, both the convective and diffusive contributions increase with shear in the buoyancy-dominant regime. Consequently, the total heat flux increases with $\varOmega$, implying that the shear from rough walls enhances heat transport.
In order to provide a clear visualization of the heat transport enhancement in the buoyancy-dominant regime, figure \ref{decompose_Nu} ($ b-e $) presents the instantaneous temperature field superposed with instantaneous velocity vectors at various non-dimensional angular velocity differences ($\varOmega$) for $Ra = 10^7$. It can be observed intuitively that the cold and hot plumes, which serve as the main heat carriers in turbulent convective heat transfer \citep{shang2003measured}, become thicker and stronger as $\varOmega$ increases. Furthermore, the convection intensifies with higher $\varOmega$, indicating an enhanced efficiency of heat transport. Additionally, the rotating rough walls function like conveyor belts, facilitating enhanced interactions between the convection rolls and secondary flows within the cavities formed by the roughness elements. This efficient pumping mechanism effectively extracts the trapped hot and cold fluids from these cavities, leading to the thinning of thermal boundary layers and intensifying the convective process (as seen in figure \ref{enlarge}). Consequently, the heat transport enhancement in this buoyancy-dominant regime is evident.

\begin{figure}
	\centering
	{\includegraphics[width=0.95\textwidth]{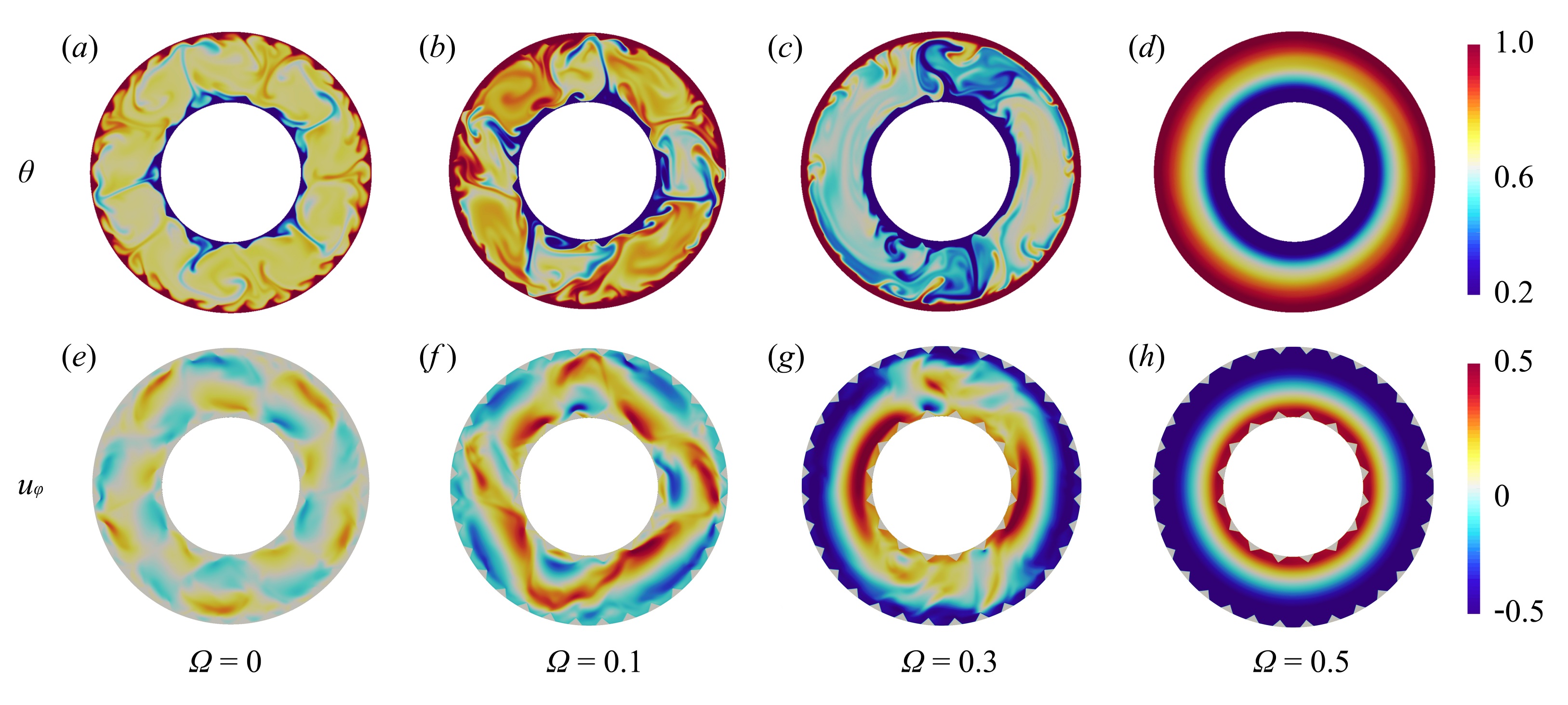}}
	\caption{The instantaneous temperature ($a-d$) and azimuthal velocity ($e-h$) fields for Case A at different non-dimensional angular velocity difference $\varOmega$ and $Ra = 10^7$.}\label{snap_tem_azi}
\end{figure}

To delve deeper into the underlying dynamics governing heat transfer behavior, our focus shifts to the evolution of flow structures under the influence of wall shear. Figure \ref{snap_tem_azi} visually presents the instantaneous temperature ($a-d$) and azimuthal velocity ($e-h$) fields for Case A at different non-dimensional angular velocity difference ($\varOmega$) for $Ra = 10^7$. According to Jiang et al. \cite{jiang2020supergravitational}, the convection rolls observed in ACRBC exhibit a prograde rotation around the axis, with a rotation rate faster than that of the experimental system. This phenomenon, known as zonal flow, has been extensively studied in experiments involving astrophysical and geophysical flows \citep{heimpel2005simulation,von2015generation}. The emergence of zonal flow can be attributed to the special geometric structure of the ACRBC. Under the influence of Coriolis force, the hot plumes detach from the outer cylinder and the cold plumes detach from the inner cylinder to deflect towards their right-hand side from their initial direction when the system rotates counterclockwise. As a result, two similar deflection angles are formed. However, due to the different curvatures of the cylinders, the similar deflection angles of the hot and cold plumes have distinct effects. The hot plumes directly impact the region where the cold plumes are ejected, leading to the anticlockwise rotation of the cold plumes and the overall flow. On the other hand, the relatively large distance between the ejection position of the hot plumes and the cold plumes implies that the motion of the hot plumes is not directly influenced by the cold plumes. Consequently, the hot plumes prevail and drive the overall flow to move in the same direction as the system rotates. Further detailed analysis can be found in our previous studies \citep{jiang2020supergravitational,wang2022effects}. It is noted that our previous research \citep{xu2024effects} on ACRBC with rough walls demonstrates that strong Coriolis forces ($Ro^{-1} = 20$, the same value used in the present study) significantly weaken the zonal flow. Additionally, we found that roughness on the outer wall promotes the formation of zonal flow, while roughness on the inner wall reduces its intensity.

In the absence of shear ($\varOmega = 0$), the snapshots of temperature and azimuthal velocity reveal the formation of five pairs of convection rolls (identified through five cold plumes detached from the inner wall in figure \ref{snap_tem_azi}\textit{a}). However, in the case of smooth ACRBC reported by Zhong et al. \cite{zhong2023sheared}, only four pairs of convection rolls are observed. As the value of $\varOmega$ increases to 0.1, both our study and the study conducted by Zhong et al. \cite{zhong2023sheared} exhibit a reduction in the number of convection rolls. At $\varOmega=0.1$, our study shows the presence of four pairs of convection rolls, whereas Zhong et al. \cite{zhong2023sheared} observes only two pairs.
In turbulent thermal convection, the convective heat transport is primarily characterized by the multi-scale interactions between convection rolls and thermal plumes detached from the boundary layers \citep{grossmann2000scaling}. The decrease in the number of convection rolls is expected to diminish the efficiency of heat transport, as observed in the smooth sheared ACRBC study conducted by Zhong et al. \cite{zhong2023sheared}. However, in our study, despite the reduction in the number of convection rolls from five to four due to wall roughness shear at $\varOmega = 0.1$, there is a significant enhancement in heat transport (see figure \ref{NuvsOmega_1}$b$). Meaning that the shear forces generated by the rotation of the rough walls are crucial to the heat transfer process. Specifically, these shear forces disturb the fluid flow over the surface, making it easier for the fluid within the cavities between the rough elements to be drawn into the flow field. By examining figure \ref{snap_tem_azi} ($b$) and ($f$), we observe that although the number of convection rolls decreases due to the shear of the rough walls, the strength of convection is intensified by the shear. This finding has also been evidenced in figure \ref{decompose_Nu}. As mentioned earlier, the rotating rough walls also act as conveyor belts, facilitating the thinning and disruption of the thermal boundary layer within the cavities. As the thermal boundary layer becomes thinner, heat is transferred more efficiently from the heated surface to the fluid, significantly improving heat transfer. In this way, the shear forces from the rotating rough walls induce turbulence, reducing the boundary layer thickness and promoting faster heat transfer, ultimately enhancing the overall heat transfer performance. Consequently, the heat transport is remarkably enhanced, approximately 2.79 times greater than the heat transport at $\varOmega = 0$.
It is important to clarify whether the enhancement of heat transport at $\varOmega = 0.1$ is caused by the zonal flow or by the rotation of rough walls. As mentioned above, roughness on the outer wall promotes the formation of zonal flow, while roughness on the inner wall reduces its intensity \citep{xu2024effects}. If the highest heat transfer efficiency at $\varOmega = 0.1$ were indeed due to enhanced zonal flow, we would expect Case C (with a rough outer wall and a smooth inner wall) to exhibit the highest $Nu$ value. However, the results presented in figure \ref{NuvsOmega} show that Case A (with both inner and outer walls rough) has the highest $Nu$ value at $\varOmega = 0.1$, suggesting that the heat transfer enhancement is not primarily driven by the zonal flow.

Upon further increasing $\varOmega$ to 0.3, the observation reveals the presence of two pairs of convection rolls, with each pair consisting of a large roll covering nearly half of the annulus and a smaller roll. It can be speculated that the shear imposed by the two boundaries stretches the plumes and enhances their azimuthal motion in the shear direction. This phenomenon of stretched plumes under shear has also been observed in sheared ACRBC \citep{zhong2023sheared} and sheared RB convection \citep{goluskin2014convectively, blass2020flow} with smooth walls. The flow motion induced by shear aligns with the motion induced by the Coriolis force, causing hot plumes to turn anticlockwise near the inner cylinder and cold plumes to turn clockwise near the outer cylinder. Consequently, the imposed shear increases the size of the large roll. Due to the system size is limited, the number of roll pairs decreases and cannot accommodate four pairs. Nevertheless, at $\varOmega = 0.3$, the turbulent strength remains higher compared to $\varOmega = 0$. As a result, the heat transport is reduced approximately 44\% compared to that at $\varOmega = 0.1$, but still exhibits an increase of approximately 93\% compared to that at $\varOmega = 0$.
Further increasing $\varOmega$ to 0.5 leads to a flow that is close to laminar and non-vortical, causing the disappearance of plumes. At this stage, convection and heat transfer are significantly suppressed, and heat transport relies mainly on conduction. The presence of plumes, which are essential heat carriers, becomes challenging. Consequently, the Nusselt number (\textit{Nu}) experiences a significant drop, reaching approximately 12\% of the value at $\varOmega = 0$. The findings from figure \ref{snap_tem_azi} indicate that the modulation effect of the rough walls on flow structures is governed by the interaction between wall shear and buoyancy, which is amply discussed in figure \ref{NuvsOmega}.

\begin{figure}
	\centering
	{\includegraphics[width=0.95\textwidth]{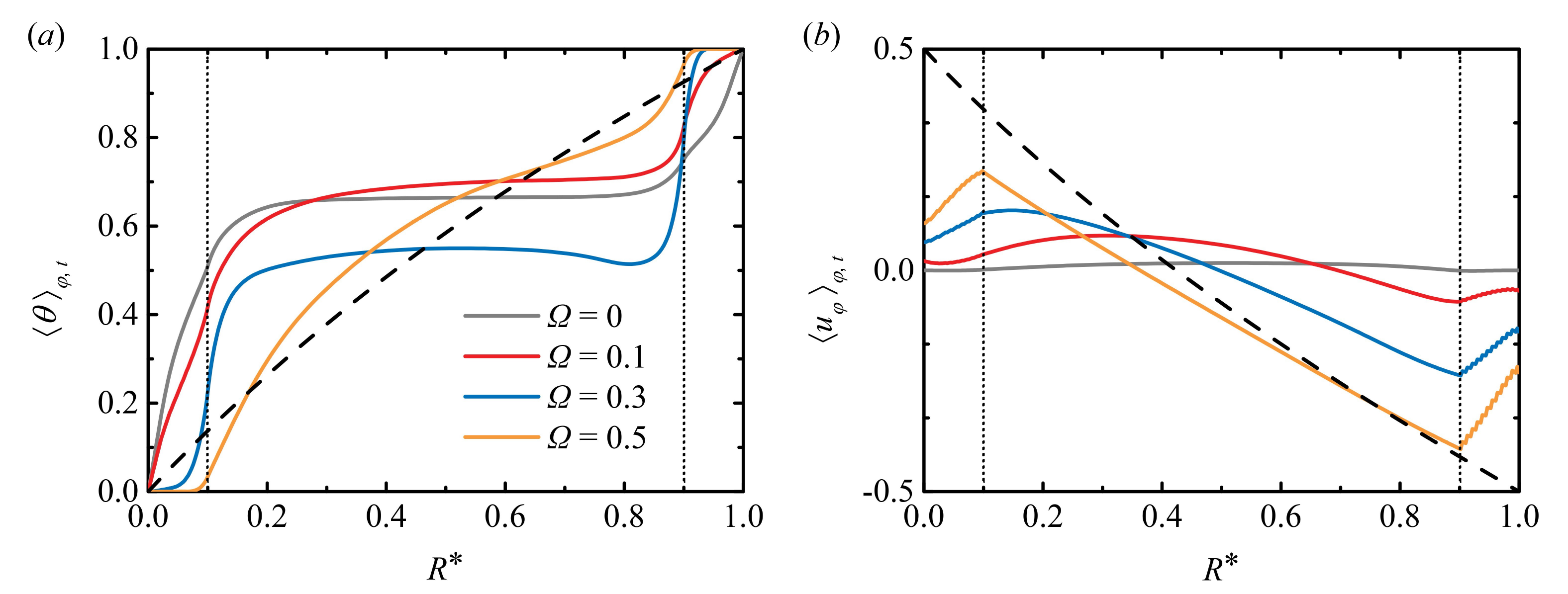}}
	\caption{Radial distribution of azimuthal and time-averaged (\textit{a}) temperature and (\textit{b}) azimuthal velocity, for Case A at different non-dimensional angular velocity difference $\varOmega$ and $Ra = 10^7$. The dashed lines represent the profiles of the laminar and non-vortical flow. The dotted lines donate the tips of roughness elements.}\label{profile_tem_azi}
\end{figure}

Figure \ref{profile_tem_azi} displays the averaged temperature and azimuthal velocity profiles for Case A at different values of $\varOmega$. The dashed lines represent the profiles for laminar and non-vortical flow. The temperature profile is described by the function $\theta = \ln(r/R_i)/\ln(R_o/R_i)$, and the azimuthal velocity profile is given by $u_\varphi = -(1+\eta^2)\varOmega r/(1-\eta^2) + 2R_i^2\varOmega/[(1-\eta^2)r]$, as proposed by Ali et al. \cite{ali1990stability} and Zhong et al. \cite{zhong2023sheared}. In figure \ref{profile_tem_azi}(\textit{a}), it can be observed that the bulk temperature deviates from the arithmetic mean temperature of the two boundaries ($\theta_m = 0.5$) when $\varOmega \le 0.1$. This asymmetry in temperature profiles without shear and with slight shear is attributed to the inherent asymmetry caused by the effects of radially dependent centrifugal buoyancy, as reported by Wang et al. \cite{wang2022effects} and Zhong et al. \cite{zhong2023sheared}. However, as the shear increases to $\varOmega = 0.3$, the bulk temperature approaches the arithmetic mean temperature ($\theta_m = 0.5$). This indicates the presence of more cold plumes generated and detached from the inner rough walls, as clearly observed in figure \ref{snap_tem_azi}(\textit{c}). With further increasing $\varOmega$, the temperature profile gradually evolves towards a laminar and non-vortical flow profile, indicating the dominance of shear at this stage.

Regarding the azimuthal velocity profile shown in \ref{profile_tem_azi}(\textit{b}), a similar behavior to the bulk temperature is not observed. At $\varOmega = 0$, the strong Coriolis force ($Ro^{-1} = 20$) effectively restricts the flow movement in the direction of system rotation, resulting in nearly zero azimuthal velocity in the absence of shear. As shear is imposed, the averaged azimuthal velocity varies from $\varOmega$ at the inner cylinder to $-\varOmega$ at the outer cylinder. As $\varOmega$ increases, the azimuthal velocity profile gradually approaches a laminar flow profile as well. It should be noted that even at the strong shear $\varOmega = 0.5$, the azimuthal velocity profile still deviates from the laminar flow profile. This can be attributed to two reasons: first, the calculated laminar flow profile assumes a radius ratio of $\eta = 0.5$, while the effective radius ratio for Case A is approximately $\eta \approx 0.58$; second, the flow is not entirely in a laminar state at this $\varOmega$ and $Ra$, as evidenced by the decreasing Nusselt number \textit{Nu} with further increasing $\varOmega$ until $\varOmega \geq 0.6$ (as shown in figure \ref{NuvsOmega}). Figure \ref{profile_tem_azi}(\textit{b}) also shows that the azimuthal velocity (which indicates the strength of the zonal flow) at $\varOmega = 0.1$ is lower compared to $\varOmega = 0.5$ and $\varOmega = 0.5$, yet the $Nu$ value at $\varOmega = 0.1$ remains the highest (as presented in figure \ref{NuvsOmega}$b$). This further supports the conclusion that the heat transfer enhancement is not caused by the strengthening of the zonal flow. Additionally, the absolute value of azimuthal velocity decreases from the top position of the roughness elements to the valley, particularly evident for large $\varOmega$, indicating that the flow within the roughness elements remains relatively stagnant compared to the boundaries. These quantitative results provide additional insight into the flow structures affected by rough walls shear and strongly support the conclusions regarding regime transition illustrated in figure \ref{NuvsOmega}.

In summary, the investigation of sheared (ACRBC) with rough walls reveals a remarkable enhancement in heat transfer efficiency, particularly in the buoyancy-dominant regime. This is in stark contrast to the cases of smooth-wall ACRBC, Couette-RBC, and Poiseuille-RBC, where the Nusselt number \textit{Nu} is initially suppressed due to shear effects. By decomposing the \textit{Nu}, it is found that both convective and diffusive contributions increase with shear, resulting in an overall rise in heat flux, indicating that shear of rough walls enhances heat transfer. Moreover, the analysis of flow structures demonstrates that as shear strength increases, the plumes become stronger, leading to intensified convection and improved heat transfer efficiency. The rough walls also act as conveyors, facilitating the interaction between convective vortices and secondary flows within the cavities formed by the roughness elements, further enhancing the convective process. As shear strength continues to rise, although the number of convective vortices decreases, their intensity increases, resulting in higher heat transfer efficiency compared to cases without shear. However, as shear further increases, the flow approaches a non-vortical laminar regime, the plumes disappear, and heat transfer becomes primarily dependent on thermal conduction, leading to a significant decrease in the Nusselt number. These findings reveal the significant impact of rough walls on flow structures and heat transfer efficiency, providing an explanation for the enhanced heat transfer observed in sheared ACRBC with rough walls.

\section{Conclusions}
In this study, we conducted extensive two-dimensional direct numerical simulations to explore the coupling effects of wall shear and roughness on annular centrifugal Rayleigh-B\'{e}nard convection (ACRBC). We considered three specific combinations of rough walls, each characterized by a uniform roughness element height of $\delta = 0.1L$. In Case A, the inner cylinder is equipped with 16 isosceles right triangles that are evenly distributed in the azimuthal direction, while the outer cylinder features 32 triangles. In Case B, the same number of roughness elements as in Case A is maintained on the inner cylinder, but the outer wall remains smooth. Conversely, Case C involves a smooth inner cylinder combined with a rough outer cylinder, where the number of roughness elements matches that of Case A. The main findings of this study can be summarized as follows:

As the non-dimensional angular velocity difference $\varOmega$ increases, the influence of different combinations of roughness follows a trend of this order (Case A $ > $ Case B $ > $ Case C) at the same $\varOmega$. By considering the critical non-dimensional angular velocity differences $\varOmega_{c1}$ and $\varOmega_{c2}$, we classified three distinct flow regimes: buoyancy-dominant, transitional, and shear-dominant. In the buoyancy-dominant regime, the flow structures resemble stretched convective rolls similar to those observed in ACRRB without shear. As the shear increases, the number of convection rolls decreases. However, the presence of rotating rough walls acts as conveyor belts, promoting interactions between the convection rolls and secondary flows within the cavities formed by the roughness elements. This efficient pumping mechanism facilitates the extraction of trapped hot and cold fluids from the cavities, resulting in the thinning of thermal boundary layers and the strengthening of convection. Consequently, heat transport (Nusselt number) is enhanced in this regime. When the non-dimensional angular velocity difference is close to the critical value $\varOmega_{c1}$, a sharp and abrupt transition occurs, leading to a significant decrease in the corresponding Nusselt number. In this transitional regime, the number of convection rolls decreases as the shear increases, eventually reaching a point where they become difficult to sustain. Beyond the critical value $\varOmega_{c2}$ of the non-dimensional angular velocity difference, the flow becomes shear-dominant. In this regime, the flow exhibits a laminar non-vortical pattern for velocity and pure conduction for temperature.
These findings have important implications for controlling heat transfer and flow dynamics in rapidly rotating machines. The understanding of the coupling effects between wall shear and roughness in ACRBC can contribute to the optimization and design of such systems.

\begin{acknowledgments}
	This study is financially supported by National Natural Science Foundation of China (11988102), Science Foundation of China University of Petroleum, Beijing (No.2462024YJRC008) and the New Cornerstone Science Foundation through the XPLORER prize.
\end{acknowledgments}

\appendix

\section{Simulation details.}\label{appendix}

\begin{table}
	\caption{Simulation parameters for Case A. The columns from left to right indicate the following: the Rayleigh number \textit{Ra}, the non-dimensional rotational speed difference $\varOmega$, the resolution employed, the maximum grid spacing $ \Delta $$ _{g} $ compared with the Kolmogorov scale estimated by the global criterion $ \eta_{K} $, the maximum grid spacing $ \Delta_{g} $ compared with the Batchelor scale estimated by $ \eta_{B} $, the calculated Nusselt numbers \textit{Nu} and their relative difference of two halves $ \epsilon_{Nu} $.}
	\label{table1}
	\begin{center}
		\begin{tabular}{lclcccrc}
			\hline
			\makebox[0.05\textwidth][l]{No.}&\makebox[0.16\textwidth][c]{$Ra$}&\makebox[0.05\textwidth][l]{$\varOmega$}&\makebox[0.2\textwidth][c]{$N_{\varphi}$$ \times $$N_{r}$}&\makebox[0.08\textwidth][c]{$\varDelta_{g}/\eta_{K}$}&\makebox[0.1\textwidth][c]{$\varDelta_{g}/\eta_{B}$}&\makebox[0.08\textwidth][r]{$Nu$}&\makebox[0.11\textwidth][c]{$\epsilon_{Nu}$}\\
			\hline
			1    & 1.0 $ \times $ 10$ ^{6} $ & 0    & 1024 $ \times $ 128 &  0.33 & 0.68 & 7.46 & 0.21$ \% $\\
			2    & 1.0 $ \times $ 10$ ^{6} $ & 0.01 & 1024 $ \times $ 128 &  0.33 & 0.69 & 7.96 & 0.27$ \% $\\
			3    & 1.0 $ \times $ 10$ ^{6} $ & 0.05 & 1024 $ \times $ 128 &  0.37 & 0.76 & 11.25 & 0.34$ \% $\\
			4    & 1.0 $ \times $ 10$ ^{6} $ & 0.1  & 1024 $ \times $ 128 &  0.39 & 0.80 & 13.76 & 0.44$ \% $\\
			5    & 1.0 $ \times $ 10$ ^{6} $ & 0.2  & 1024 $ \times $ 128 &  0.39 & 0.80 & 13.82 & 0.43$ \% $\\
			6    & 1.0 $ \times $ 10$ ^{6} $ & 0.3  & 1024 $ \times $ 128 &  0.35 & 0.72 & 9.14 & 0.33$ \% $\\
			7    & 1.0 $ \times $ 10$ ^{6} $ & 0.35 & 1024 $ \times $ 128 &  0.32 & 0.65 & 6.65 & 0.38$ \% $\\
			8    & 1.0 $ \times $ 10$ ^{6} $ & 0.4  & 1024 $ \times $ 128 &  0.23 & 0.48 & 2.61 & 0.45$ \% $\\
			9    & 1.0 $ \times $ 10$ ^{6} $ & 0.5  & 1024 $ \times $ 128 &  0.17 & 0.35 & 1.45 & 0.41$ \% $\\
			10   & 1.0 $ \times $ 10$ ^{6} $ & 0.6  & 1024 $ \times $ 128 &  0.15 & 0.30 & 1.26 & 0.54$ \% $\\
			11   & 1.0 $ \times $ 10$ ^{6} $ & 1    & 1024 $ \times $ 128 &  0.14 & 0.30 & 1.25 & 0.62$ \% $\\
			12   & 1.0 $ \times $ 10$ ^{7} $ & 0    & 1536 $ \times $ 192 &  0.27 & 0.56 & 16.23 & 0.13$ \% $\\
			13   & 1.0 $ \times $ 10$ ^{7} $ & 0.01 & 1536 $ \times $ 192 &  0.28 & 0.58 & 18.33 & 0.21$ \% $\\
			14   & 1.0 $ \times $ 10$ ^{7} $ & 0.05 & 1536 $ \times $ 192 &  0.32 & 0.67 & 33.01 & 0.39$ \% $\\
			15   & 1.0 $ \times $ 10$ ^{7} $ & 0.1  & 1536 $ \times $ 192 &  0.35 & 0.73 & 45.21 & 0.28$ \% $\\
			16   & 1.0 $ \times $ 10$ ^{7} $ & 0.2  & 1536 $ \times $ 192 &  0.36 & 0.75 & 49.35 & 0.32$ \% $\\
			17   & 1.0 $ \times $ 10$ ^{7} $ & 0.3  & 1536 $ \times $ 192 &  0.32 & 0.66 & 31.31 & 0.36$ \% $\\
			18   & 1.0 $ \times $ 10$ ^{7} $ & 0.4  & 1536 $ \times $ 192 &  0.29 & 0.60 & 20.94 & 0.35$ \% $\\
			19   & 1.0 $ \times $ 10$ ^{7} $ & 0.45 & 1536 $ \times $ 192 &  0.24 & 0.49 & 9.87 & 0.48$ \% $\\
			20   & 1.0 $ \times $ 10$ ^{7} $ & 0.5  & 1536 $ \times $ 192 &  0.15 & 0.32 & 2.64 & 0.51$ \% $\\
			21   & 1.0 $ \times $ 10$ ^{7} $ & 0.6  & 1536 $ \times $ 192 &  0.10 & 0.20 & 1.27 & 0.59$ \% $\\
			22   & 1.0 $ \times $ 10$ ^{7} $ & 1    & 1536 $ \times $ 192 &  0.10 & 0.20 & 1.26 & 0.60$ \% $\\
			23   & 1.0 $ \times $ 10$ ^{8} $ & 0    & 2048 $ \times $ 256 &  0.45 & 0.93 & 37.13 & 0.42$ \% $\\
			24   & 1.0 $ \times $ 10$ ^{8} $ & 0.01 & 2048 $ \times $ 256 &  0.50 & 1.03 & 56.78 & 0.33$ \% $\\
			25   & 1.0 $ \times $ 10$ ^{8} $ & 0.05 & 2048 $ \times $ 256 &  0.60 & 1.24 & 118.50 & 0.29$ \% $\\
			26   & 1.0 $ \times $ 10$ ^{8} $ & 0.1  & 2048 $ \times $ 256 &  0.68 & 1.40 & 191.33 & 0.37$ \% $\\
			27   & 1.0 $ \times $ 10$ ^{8} $ & 0.2  & 2048 $ \times $ 256 &  0.73 & 1.51 & 258.88 & 0.26$ \% $\\
			28   & 1.0 $ \times $ 10$ ^{8} $ & 0.3  & 2048 $ \times $ 256 &  0.66 & 1.38 & 177.51 & 0.43$ \% $\\
			29   & 1.0 $ \times $ 10$ ^{8} $ & 0.4  & 2048 $ \times $ 256 &  0.56 & 1.15 & 88.12 & 0.38$ \% $\\
			30   & 1.0 $ \times $ 10$ ^{8} $ & 0.5  & 2048 $ \times $ 256 &  0.48 & 0.99 & 48.38 & 0.43$ \% $\\
			31   & 1.0 $ \times $ 10$ ^{8} $ & 0.6  & 2048 $ \times $ 256 &  0.22 & 0.45 & 3.05 & 0.60$ \% $\\
			32   & 1.0 $ \times $ 10$ ^{8} $ & 0.7  & 2048 $ \times $ 256 &  0.13 & 0.27 & 1.28 & 0.66$ \% $\\
			33   & 1.0 $ \times $ 10$ ^{8} $ & 1    & 2048 $ \times $ 256 &  0.13 & 0.27 & 1.28 & 0.59$ \% $\\
			\hline
		\end{tabular}
	\end{center}
\end{table}

\begin{table}
	\caption{Simulation parameters for Case B.}
	\label{table2}
	\begin{center}
		\begin{tabular}{lclcccrc}
			\hline
			\makebox[0.05\textwidth][l]{No.}&\makebox[0.16\textwidth][c]{$Ra$}&\makebox[0.05\textwidth][l]{$\varOmega$}&\makebox[0.2\textwidth][c]{$N_{\varphi}$$ \times $$N_{r}$}&\makebox[0.08\textwidth][c]{$\varDelta_{g}/\eta_{K}$}&\makebox[0.1\textwidth][c]{$\varDelta_{g}/\eta_{B}$}&\makebox[0.08\textwidth][r]{$Nu$}&\makebox[0.11\textwidth][c]{$\epsilon_{Nu}$}\\
			\hline
			1    & 1.0 $ \times $ 10$ ^{6} $ & 0    & 1024 $ \times $ 128 &  0.32 & 0.67 & 7.32 & 0.28$ \% $\\
			2    & 1.0 $ \times $ 10$ ^{6} $ & 0.01 & 1024 $ \times $ 128 &  0.33 & 0.68 & 7.56 & 0.24$ \% $\\
			3    & 1.0 $ \times $ 10$ ^{6} $ & 0.05 & 1024 $ \times $ 128 &  0.34 & 0.71 & 8.76 & 0.31$ \% $\\
			4    & 1.0 $ \times $ 10$ ^{6} $ & 0.1  & 1024 $ \times $ 128 &  0.36 & 0.74 & 10.24 & 0.35$ \% $\\
			5    & 1.0 $ \times $ 10$ ^{6} $ & 0.2  & 1024 $ \times $ 128 &  0.35 & 0.73 & 9.77 & 0.37$ \% $\\
			6    & 1.0 $ \times $ 10$ ^{6} $ & 0.3  & 1024 $ \times $ 128 &  0.33 & 0.69 & 8.13 & 0.47$ \% $\\
			7    & 1.0 $ \times $ 10$ ^{6} $ & 0.35 & 1024 $ \times $ 128 &  0.27 & 0.56 & 3.94 & 0.27$ \% $\\
			8    & 1.0 $ \times $ 10$ ^{6} $ & 0.4  & 1024 $ \times $ 128 &  0.22 & 0.46 & 2.43 & 0.36$ \% $\\
			9    & 1.0 $ \times $ 10$ ^{6} $ & 0.5  & 1024 $ \times $ 128 &  0.16 & 0.33 & 1.38 & 0.53$ \% $\\
			10   & 1.0 $ \times $ 10$ ^{6} $ & 0.6  & 1024 $ \times $ 128 &  0.12 & 0.26 & 1.13 & 0.46$ \% $\\
			11   & 1.0 $ \times $ 10$ ^{6} $ & 1    & 1024 $ \times $ 128 &  0.12 & 0.26 & 1.13 & 0.57$ \% $\\
			12   & 1.0 $ \times $ 10$ ^{7} $ & 0    & 1536 $ \times $ 192 &  0.26 & 0.55 & 15.05 & 0.26$ \% $\\
			13   & 1.0 $ \times $ 10$ ^{7} $ & 0.01 & 1536 $ \times $ 192 &  0.27 & 0.56 & 16.11 & 0.44$ \% $\\
			14   & 1.0 $ \times $ 10$ ^{7} $ & 0.05 & 1536 $ \times $ 192 &  0.29 & 0.61 & 22.35 & 0.62$ \% $\\
			15   & 1.0 $ \times $ 10$ ^{7} $ & 0.1  & 1536 $ \times $ 192 &  0.30 & 0.62 & 24.47 & 0.61$ \% $\\
			16   & 1.0 $ \times $ 10$ ^{7} $ & 0.2  & 1536 $ \times $ 192 &  0.30 & 0.62 & 23.41 & 0.45$ \% $\\
			17   & 1.0 $ \times $ 10$ ^{7} $ & 0.3  & 1536 $ \times $ 192 &  0.28 & 0.58 & 18.17 & 0.48$ \% $\\
			18   & 1.0 $ \times $ 10$ ^{7} $ & 0.4  & 1536 $ \times $ 192 &  0.25 & 0.52 & 11.96 & 0.51$ \% $\\
			19   & 1.0 $ \times $ 10$ ^{7} $ & 0.45 & 1536 $ \times $ 192 &  0.20 & 0.41 & 5.54 & 0.57$ \% $\\
			20   & 1.0 $ \times $ 10$ ^{7} $ & 0.5  & 1536 $ \times $ 192 &  0.15 & 0.32 & 2.59 & 0.43$ \% $\\
			21   & 1.0 $ \times $ 10$ ^{7} $ & 0.6  & 1536 $ \times $ 192 &  0.08 & 0.17 & 1.14 & 0.48$ \% $\\
			22   & 1.0 $ \times $ 10$ ^{7} $ & 1    & 1536 $ \times $ 192 &  0.08 & 0.17 & 1.13 & 0.65$ \% $\\
			23   & 1.0 $ \times $ 10$ ^{8} $ & 0    & 2048 $ \times $ 256 &  0.43 & 0.89 & 31.42 & 0.39$ \% $\\
			24   & 1.0 $ \times $ 10$ ^{8} $ & 0.01 & 2048 $ \times $ 256 &  0.45 & 0.92 & 36.75 & 0.41$ \% $\\
			25   & 1.0 $ \times $ 10$ ^{8} $ & 0.05 & 2048 $ \times $ 256 &  0.48 & 0.99 & 47.83 & 0.38$ \% $\\
			26   & 1.0 $ \times $ 10$ ^{8} $ & 0.1  & 2048 $ \times $ 256 &  0.48 & 1.00 & 50.67 & 0.42$ \% $\\
			27   & 1.0 $ \times $ 10$ ^{8} $ & 0.2  & 2048 $ \times $ 256 &  0.47 & 0.97 & 45.27 & 0.43$ \% $\\
			28   & 1.0 $ \times $ 10$ ^{8} $ & 0.3  & 2048 $ \times $ 256 &  0.45 & 0.93 & 37.94 & 0.56$ \% $\\
			29   & 1.0 $ \times $ 10$ ^{8} $ & 0.4  & 2048 $ \times $ 256 &  0.44 & 0.92 & 35.78 & 0.49$ \% $\\
			30   & 1.0 $ \times $ 10$ ^{8} $ & 0.5  & 2048 $ \times $ 256 &  0.43 & 0.88 & 31.01 & 0.52$ \% $\\
			31   & 1.0 $ \times $ 10$ ^{8} $ & 0.6  & 2048 $ \times $ 256 &  0.20 & 0.42 & 2.57 & 0.48$ \% $\\
			32   & 1.0 $ \times $ 10$ ^{8} $ & 0.7  & 2048 $ \times $ 256 &  0.11 & 0.24 & 1.15 & 0.55$ \% $\\
			33   & 1.0 $ \times $ 10$ ^{8} $ & 1    & 2048 $ \times $ 256 &  0.11 & 0.23 & 1.14 & 0.64$ \% $\\
			\hline
		\end{tabular}
	\end{center}
\end{table}

\begin{table}
	\caption{Simulation parameters for Case C.}
	\label{table3}
	\begin{center}
		\begin{tabular}{lclcccrc}
			\hline
			\makebox[0.05\textwidth][l]{No.}&\makebox[0.16\textwidth][c]{$Ra$}&\makebox[0.05\textwidth][l]{$\varOmega$}&\makebox[0.2\textwidth][c]{$N_{\varphi}$$ \times $$N_{r}$}&\makebox[0.08\textwidth][c]{$\varDelta_{g}/\eta_{K}$}&\makebox[0.1\textwidth][c]{$\varDelta_{g}/\eta_{B}$}&\makebox[0.08\textwidth][r]{$Nu$}&\makebox[0.11\textwidth][c]{$\epsilon_{Nu}$}\\
			\hline
			1    & 1.0 $ \times $ 10$ ^{6} $ & 0    & 1024 $ \times $ 128 &  0.32 & 0.67 & 7.32 & 0.32$ \% $\\
			2    & 1.0 $ \times $ 10$ ^{6} $ & 0.01 & 1024 $ \times $ 128 &  0.33 & 0.68 & 7.44 & 0.19$ \% $\\
			3    & 1.0 $ \times $ 10$ ^{6} $ & 0.05 & 1024 $ \times $ 128 &  0.33 & 0.69 & 8.03 & 0.28$ \% $\\
			4    & 1.0 $ \times $ 10$ ^{6} $ & 0.1  & 1024 $ \times $ 128 &  0.34 & 0.70 & 8.43 & 0.36$ \% $\\
			5    & 1.0 $ \times $ 10$ ^{6} $ & 0.2  & 1024 $ \times $ 128 &  0.32 & 0.66 & 6.76 & 0.38$ \% $\\
			6    & 1.0 $ \times $ 10$ ^{6} $ & 0.3  & 1024 $ \times $ 128 &  0.30 & 0.62 & 5.59 & 0.45$ \% $\\
			7    & 1.0 $ \times $ 10$ ^{6} $ & 0.35 & 1024 $ \times $ 128 &  0.25 & 0.52 & 3.28 & 0.47$ \% $\\
			8    & 1.0 $ \times $ 10$ ^{6} $ & 0.4  & 1024 $ \times $ 128 &  0.22 & 0.46 & 2.41 & 0.32$ \% $\\
			9    & 1.0 $ \times $ 10$ ^{6} $ & 0.5  & 1024 $ \times $ 128 &  0.16 & 0.33 & 1.35 & 0.49$ \% $\\
			10   & 1.0 $ \times $ 10$ ^{6} $ & 0.6  & 1024 $ \times $ 128 &  0.11 & 0.22 & 1.07 & 0.51$ \% $\\
			11   & 1.0 $ \times $ 10$ ^{6} $ & 1    & 1024 $ \times $ 128 &  0.11 & 0.22 & 1.07 & 0.53$ \% $\\
			12   & 1.0 $ \times $ 10$ ^{7} $ & 0    & 1536 $ \times $ 192 &  0.26 & 0.54 & 14.03 & 0.22$ \% $\\
			13   & 1.0 $ \times $ 10$ ^{7} $ & 0.01 & 1536 $ \times $ 192 &  0.27 & 0.55 & 15.30 & 0.32$ \% $\\
			14   & 1.0 $ \times $ 10$ ^{7} $ & 0.05 & 1536 $ \times $ 192 &  0.28 & 0.57 & 17.46 & 0.38$ \% $\\
			15   & 1.0 $ \times $ 10$ ^{7} $ & 0.1  & 1536 $ \times $ 192 &  0.28 & 0.57 & 17.92 & 0.45$ \% $\\
			16   & 1.0 $ \times $ 10$ ^{7} $ & 0.2  & 1536 $ \times $ 192 &  0.26 & 0.53 & 13.69 & 0.51$ \% $\\
			17   & 1.0 $ \times $ 10$ ^{7} $ & 0.3  & 1536 $ \times $ 192 &  0.24 & 0.50 & 10.69 & 0.44$ \% $\\
			18   & 1.0 $ \times $ 10$ ^{7} $ & 0.4  & 1536 $ \times $ 192 &  0.22 & 0.46 & 7.67 & 0.29$ \% $\\
			19   & 1.0 $ \times $ 10$ ^{7} $ & 0.45 & 1536 $ \times $ 192 &  0.17 & 0.35 & 3.24 & 0.37$ \% $\\
			20   & 1.0 $ \times $ 10$ ^{7} $ & 0.5  & 1536 $ \times $ 192 &  0.15 & 0.31 & 2.38 & 0.33$ \% $\\
			21   & 1.0 $ \times $ 10$ ^{7} $ & 0.6  & 1536 $ \times $ 192 &  0.07 & 0.15 & 1.08 & 0.26$ \% $\\
			22   & 1.0 $ \times $ 10$ ^{7} $ & 1    & 1536 $ \times $ 192 &  0.07 & 0.15 & 1.07 & 0.44$ \% $\\
			23   & 1.0 $ \times $ 10$ ^{8} $ & 0    & 2048 $ \times $ 256 &  0.42 & 0.87 & 28.81 & 0.51$ \% $\\
			24   & 1.0 $ \times $ 10$ ^{8} $ & 0.01 & 2048 $ \times $ 256 &  0.43 & 0.90 & 32.88 & 0.41$ \% $\\
			25   & 1.0 $ \times $ 10$ ^{8} $ & 0.05 & 2048 $ \times $ 256 &  0.43 & 0.90 & 32.98 & 0.32$ \% $\\
			26   & 1.0 $ \times $ 10$ ^{8} $ & 0.1  & 2048 $ \times $ 256 &  0.43 & 0.90 & 33.49 & 0.43$ \% $\\
			27   & 1.0 $ \times $ 10$ ^{8} $ & 0.2  & 2048 $ \times $ 256 &  0.41 & 0.85 & 26.50 & 0.34$ \% $\\
			28   & 1.0 $ \times $ 10$ ^{8} $ & 0.3  & 2048 $ \times $ 256 &  0.38 & 0.78 & 19.05 & 0.51$ \% $\\
			29   & 1.0 $ \times $ 10$ ^{8} $ & 0.4  & 2048 $ \times $ 256 &  0.37 & 0.78 & 18.78 & 0.52$ \% $\\
			30   & 1.0 $ \times $ 10$ ^{8} $ & 0.5  & 2048 $ \times $ 256 &  0.33 & 0.68 & 11.47 & 0.34$ \% $\\
			31   & 1.0 $ \times $ 10$ ^{8} $ & 0.6  & 2048 $ \times $ 256 &  0.17 & 0.33 & 1.55 & 0.49$ \% $\\
			32   & 1.0 $ \times $ 10$ ^{8} $ & 0.7  & 2048 $ \times $ 256 &  0.10 & 0.21 & 1.09 & 0.57$ \% $\\
			33   & 1.0 $ \times $ 10$ ^{8} $ & 1    & 2048 $ \times $ 256 &  0.10 & 0.20 & 1.08 & 0.61$ \% $\\
			\hline
		\end{tabular}
	\end{center}
\end{table}

\end{document}